\documentclass[twocolumn,aps,prx,groupedaddress]{revtex4}
\usepackage{graphicx,wrapfig,amsmath,amssymb,enumitem,upgreek,mathtools,amsfonts,physics,multirow,color}
\usepackage{hyperref}
\usepackage[normalem]{ulem}

\newcommand{\ov}[1]{\overline{#1}}

\renewcommand{\th}{\vartheta}

\renewcommand{\Re}{{\rm \, Re\,}}
\renewcommand{\Im}{{\rm \, Im\,}}

\renewcommand{\vec}[1]{{\boldsymbol #1}}

\def\bk{{\vec k}}
\def\bq{{\vec q}}

\begin{document}
\title{Fractional Chern Insulator States in Twisted Bilayer Graphene: An Analytical Approach}
\author{Patrick J. Ledwith, Grigory Tarnopolsky, Eslam Khalaf and Ashvin Vishwanath}
\affiliation{Department of Physics,  Harvard University,
Cambridge, MA 02138, USA}

\begin{abstract}
  Recent experiments  on  bilayer graphene  twisted near the magic angle have observed spontaneous integer quantum Hall states in the presence of an aligned hexagonal boron nitride (hBN) substrate. These states  
  arise from the complete filling of Chern bands. A natural question is whether fractional  filling of the same bands would lead to fractional Chern insulators (FCIs) i.e. fractional quantum Hall states realized in the absence of a magnetic field.   Here, we  argue that the magic angle graphene bands have favorable quantum geometry for realizing FCIs. We show that in the  tractable `chiral' limit, the flat bands wavefunctions are an analytic function of the  crystal momentum. This remarkable property fixes the quantum metric up to an overall momentum dependent scale factor, the local Berry curvature, whose variation is itself small.
  Thus  the three  conditions associated with FCI stability (i) narrow bands (ii) `ideal' quantum metric and (iii) relatively uniform Berry curvature are all satisfied.   Our work emphasizes continuum real space approaches to FCIs in contrast to earlier works which mostly focused on tight binding models. This enables us to construct a Laughlin wavefunction on a real-space torus that is a zero energy ground state of the Coulomb interaction in the limit of very short screening length.
Finally we discuss evolution of the band geometry on tuning away from the chiral limit and show numerically that 
some of 
the desirable properties continue to hold at a quantitative level for realistic parameter values.
  
\end{abstract}
\date{\today}
\maketitle

\section{Introduction}


The recent discovery of correlated insulators and superconductors in bilayer graphene twisted near a magic angle has brought the physics of moir\'{e} superlattices to the fore.  A recent striking observation is that of a quantized Hall response in magic angle bilayer graphene. This has been attributed to the presence of an aligned hBN substrate, which breaks the $C_2$ rotation symmetry relating opposite valleys  generating bands with valley Chern number \cite{DavidGG,AndreaYoung,Zhang19,Bultinck19,XieMacDonald}. Spontaneous spin and valley polarization then leads to a Chern insulator with integer quantized Hall conductance. Similar physics is also expected in other moir\'e platforms where $C_2$ symmetry is broken such as ABC trilayer graphene on hBN \cite{FengWang,Zhang2018,FengWangDavidGGetal} and twisted double bilayer graphene \cite{TDBGexp2019, IOP_TDBG, TDBG_Pablo, Lee2019}.

 In this work we will be mainly concerned with the ground state at fractional electron filling (assuming spin and valley polarization) of the nearly flat bands of magic angle graphene with aligned hBN. Of course, in fractionally filled Landau levels (e.g. electrons in the lowest Landau level at 1/3 filling) fractional quantum Hall states arise. In fact, in the related problem of strained graphene, where Landau levels of opposite effective magnetic field arise from the opposite valleys \cite{Crommie}, a fractional filling was numerically shown to lead to spontaneous valley polarization and fractional quantum Hall states \cite{Ghaemi2012}. This is particularly relevant for twisted bilayer graphene due to the close analogy between the nearly flat bands of twisted bilayer graphene (TBG) and Landau levels which was highlighted in several recent works \cite{Tarnopolsky2019, StiefelWhitney, Po2018faithful, Song, LiuLiuDai, BultinckKhalaf}, and is all the more apparent in the presence of a hBN substrate that leads to a staggered sublattice potential and isolates Chern bands in each valley. However, the flat band wavefunctions differ crucially from the Landau levels due to the presence of actual (moir\'e) translational symmetry instead of continuous magnetic translations. As a result, the relevant problem to TBG is that of fractional Chern insulators (FCIs) which investigates the necessary ingredients for realizing the fractional quantum hall phases in systems with discrete translation symmetry.





Previous studies of FCI physics have performed numerical and theoretical analysis primarily of tight-binding models, see the reviews \cite{Parameswaran2013,Bergholtz2013}.  
Tight binding models can be related to lowest Landau level (LLL) physics by mapping the LLL Hilbert space and pseudopotential interactions to that of a lattice model \cite{Qi2011,Lee2013}. This ``ideal" lattice model will necessarily host an FCI state, and one may test adiabatic continuity between it and a realistic model \cite{Liu2013}.
However, the construction of the mapping depends on an arbitrary $\vec k$-dependent phase choice and the ideal lattice models one obtains are often unrealistic.  

While many numerical studies (see numerics discussed and cited in the comprehensive reviews \cite{Parameswaran2013,Bergholtz2013})  have successfully demonstrated that FCI states exist and agree with Chern-Simons effective field theory, microscopic theoretical analysis on FCI stability has been hampered by the fact that tight binding models lack continuous magnetic translation symmetry and rotation symmetry 
due to the discreteness of the lattice. 
For this reason, most analytic work has focused on identifying ideal  characteristics of FCI models in \emph{momentum-space}.

While flat bands and non-zero Chern number are obvious conditions for an ideal FCI state, band geometry also plays a role.  It was noticed early on that flat Berry curvature was important for realizing FCI states; for example Ref.~\onlinecite{Parameswaran2012} showed its importance for the long wavelength limit of the Girvin-MacDonald-Platzmann (GMP) algebra \cite{GMP}, a Hamiltonian formulation of the quantum Hall problem presented entirely in terms of projected density operators that obey said algebra. However, higher Landau levels do not display the same fractional quantum hall effect (FQHE) as the LLL despite having flat bands and flat Berry curvature. This suggests that other band geometric conditions involving relationships between the Fubini-Study metric and the Berry curvature play a crucial role in stabilizing FCIs by mimicking the band geometry of the LLL. Under these stricter band-geometric conditions \cite{Roy2014,Parameswaran2013}, the GMP algebra is satisfied to all orders.  One may also construct a first quantized description of Chern bands in momentum space by assuming these geometric conditions \cite{Claassen2015}. Exact diagonalization studies have shown that band geometry correlates strongly with the stability of FCI phases\cite{Jackson2015}.  Motivated by this, some models with ideal band geometry have been categorized and fine-tuned to have nearly-flat Berry curvature \cite{Lee2017}.

Here, we will show that chiral magic angle TBG (chiral-MATBG) has nearly ideal band geometry for FCI phases. Due to the unique properties of the TBG bands which we will explain in detail in this work, models like it were missed in previous discussions of ideal FCI bands. More specifically, we will show that chiral-MATBG has relatively uniform Berry curvature and satisfies the so-called isotropic ideal droplet condition (also referred to as the trace condition), discussed and defined below.

The ideal criteria above are phrased in momentum space.  Indeed, most toy FCI models are built on real space tight binding lattices whose discreteness makes geometric arguments in real space difficult.  However, we will use the continuum-nature of the model to show that a first quantized description in \emph{real} space is possible.  We will construct a Laughlin state on a real-space torus, built from chiral-MATBG wavefunctions, that is a zero-energy ground state for pseudopotential-like interactions. To our knowledge, such analysis has not appeared in the FCI literature due to the focus on tight-binding models, but see Ref. \cite{Kol1993} for the FQHE in a periodic potential.
We will also show that the chiral-MATBG wavefunctions may be interpreted as wavefunctions of a generalized LLL, in which the magnetic field is modulated in space, in the sense that density-density interactions have identical matrix elements for both systems. This provides a natural pathway to test adiabatic continuity from the LLL to potential realistic FCI states in TBG and a potential new tool for theoretical FCI analysis. 

In the next section we review the chiral model for TBG \cite{Tarnopolsky2019} to set notation and results that will be convenient for the rest of the paper.  We also discuss the introduction of a sublattice potential that separates the two flat bands so that the lower (or upper) band can host a FCI state.  If the sublattice potential is chosen to be the same on both layers the chiral-MATBG wavefunctions are unchanged but the two bands are gapped from each other.  In section \ref{bandgeometry}, we review band-geometric conditions for FCIs and show that chiral-MATBG has near-ideal band geometry; the non-ideality is inhomogeneous Berry curvature. 
In section \ref{Torus}, we construct a Laughlin ground state on a real space torus \cite{Haldane1985} which is an exact zero-energy ground state for pseudopotential-like interactions \cite{Haldane1983}.  In section \ref{curvedLLL} we show that the chiral-MATBG wavefunctions are very similar to wavefunctions for a Dirac particle in an inhomogeneous magnetic field.  Indeed, the only difference is a position dependent phase and layer polarization for the chiral-MATBG wavefunctions which does not matter for matrix elements of density operators. Finally, in section \ref{realistic} we show quantitatively using numerics how the momentum space energetics and geometry change as one moves from the chiral model to more realistic models for TBG.  The chiral model, characterized by a parameter $\kappa = w_0/w_1 = 0$, has the most promising band energetics and geometry.  As $\kappa$ increases to realistic values, $\approx 0.7-0.8$\,\cite{Koshino2018,Carr2019}, these characteristics worsen, though the chiral model bands are still adiabatically connected to realistic TBG bands.  Because FCI states are stable to small perturbations and the interaction screening length is highly tunable \cite{EfetovScreening, YoungScreening}, we argue that FCI states could be realizable in realistic models of TBG for some screening lengths.  Nonetheless, it is worth noting that if $\kappa$ can be lowered by modifying the system (it can be raised by adding dopants \cite{larson2019}) then this would likely have a positive impact on FCI stability. Pressure may have this effect because it could increase lattice relaxation.

\section{The Chiral model for TBG with a Sublattice potential} \label{chiralmodel}

The continuum model for a single spin and valley of TBG is given by the Hamiltonian \cite{Bistritzer2011,Tarnopolsky2019}
\begin{equation}
  H = \begin{pmatrix} 
    -iv\vec{\sigma}_{\theta/2} \cdot \vec{\nabla} & w T(\textbf{r}) \\ 
    w T^\dag (\textbf{r})  & -iv\vec{\sigma}_{-\theta/2} \cdot \vec{\nabla}
  \end{pmatrix},
    \label{continuummodel}
\end{equation}
which acts on a spinor $\Psi(r) = (\psi_1 ,\chi_1, \psi_2, \chi_2)$. Here, the rotated sublattice Pauli matrices are $\vec \sigma_{\theta/2} = e^{-i\theta \sigma_z/4} (\sigma_x, \sigma_y) e^{i\theta \sigma_z/4}$, and $v$ is the monolayer Fermi velocity.  The overall tunneling strength is $w$ and the functions
\begin{equation}
  \begin{aligned}
  T(\textbf{r}) & = \sum_{n=1}^{3} T_n e^{-i\textbf{q}_n \textbf{r}} ,  \\
  T_{n+1} & = \kappa \sigma_0 + \cos(n\phi)\sigma_x + \cos(n\phi)\sigma_y, 
  \end{aligned}
  \label{tunneling}
\end{equation}
characterize the different types of tunneling (AA vs AB/BA) with relative strength $\kappa$, and $\phi = 2\pi/3$.  We focus on the chiral model which corresponds to $\kappa = 0$. The vectors $\textbf{q}_n$ are given by $\textbf{q}_1 = k_\theta(0,-1)$ and $\textbf{q}_{2,3} = k_\theta(\pm \sqrt{3}/2,1/2)$, where $k_\theta = 2k_D\sin(\theta/2)$ is the moir\'{e} wavevector and $k_D = 4\pi/3a_0$ is the Dirac wavevector in terms of the monolayer lattice constant $a_0$.  

The lattice vectors and triangular lattice coordinates and reciprocal lattice analogues are
\begin{equation}
  \begin{aligned}
    \textbf{a}_{1,2} & = \frac{4\pi}{3 k_\theta} (\pm \frac{\sqrt{3}}{2}, \frac{1}{2}),  \quad
    &\textbf{b}_{1,2} 
    = k_\theta(\pm \frac{\sqrt{3}}{2},\frac{3}{2} ) \\
     r_{1,2} & = \frac{\textbf{r} \cdot \textbf{b}_{1,2}}{2\pi}, & k_{1,2}  = \frac{\textbf{k} \cdot \textbf{a}_{1,2}}{2\pi}.
  \end{aligned}
  \label{latticevecs}
\end{equation}

The chiral model ($\kappa = 0$) enjoys many simplifications. First, the $\theta$ dependence in the Pauli matrices can be removed by a gauge transformation such that $\vec \sigma_{\pm \theta/2} \to \vec \sigma$, leading to an exact particle-hole symmetry \cite{Tarnopolsky2019, BultinckKhalaf}. Second and most importantly, the single-particle Hamiltonian anticommutes with the chirality operator $\sigma_z$ which denotes the graphene sublattice. 
Additionally, the Hamiltonian can now be written in a dimensionless form $H_{\rm chiral} = E_0\mathcal{H}_{\rm chiral}$ where $E_0 = vk_\theta$ and 
\begin{equation}
  \begin{aligned}
  \mathcal{H}_{\rm chiral} & = \begin{pmatrix} 0 & D^{\dag}(\textbf{r}) \\ D(\textbf{r}) & 0 \end{pmatrix} ,  \\
  D(\textbf{r}) & = \begin{pmatrix} -k_\theta^{-1} 2i\bar \partial  & \alpha U(\textbf{r})  \\ \alpha U(-\textbf{r}) & -k_\theta^{-1}2i\bar \partial \end{pmatrix}.
\end{aligned}
  \label{chHam}
\end{equation}
The Hamiltonian now acts on the spinor $\Psi(r) = (\psi_1, \psi_2, \chi_1, \chi_2)^T$.  The anti-holomorphic derivative is given by $\bar \partial = \frac{1}{2}(\partial_x + i \partial_y)$, corresponding to the complex coordinate $z = x+iy$. The parameter $\alpha = w/E_0 \sim 1/\sin \theta$ measures the strength of tunneling.

At certain values of $\alpha$, corresponding to magic angles for $\theta$, the Hamiltonian \eqref{chHam} exhibits two exactly flat bands at zero energy.  This occurs for infinitely many $\alpha$, with a quasiperiodicity $\alpha_n \approx \alpha_{n-1} + 3/2$ \cite{Tarnopolsky2019}.  We now obtain the wavefunctions for the exactly flat bands at these values of $\alpha$ and show that they exhibit very interesting properties.

We are interested in zero modes of the Hamiltonian \eqref{chHam}.  Because the ``chirality'' matrix
  \begin{equation}
    \sigma_z = \begin{pmatrix} I_{2\times 2} & 0  \\ 0 & -I_{2 \times 2} \end{pmatrix}
    \label{chirality matrix}
  \end{equation}
  anticommutes with the Hamiltonian \eqref{chHam}, we can label the zero energy solutions with their $\sigma_z$ eigenvalue.  We therefore seek solutions to 
  \begin{equation}
    D(\textbf{r}) \psi = 0, \quad \psi(\textbf{r}) = \begin{pmatrix} \psi_1( \textbf{r}) \\ \psi_2( \textbf{r}) \end{pmatrix}
    \label{zeromodes}
\end{equation}
for the positive chirality, $\sigma_z = +1$, eigenspace.  The solutions for the negative chirality eigenspace are then $\chi( \textbf{r} ) = \psi^*(- \textbf{r})$.  

We pause to recall that the Hamiltonian \eqref{chHam} is not in traditional Bloch form; the two layers have a crystal momentum that differs by $\textbf{q}_{1}$.  The ``Bloch'' states, of positive chirality, are then of the form \begin{equation}
  \psi_{\textbf{k}}(\textbf{r}) =  e^{i  \textbf{k} \cdot \textbf{r}} \begin{pmatrix} u_{1 \textbf{k}}( \textbf{r}) \\ u_{2  \textbf{k}}( \textbf{r})e^{i\textbf{q}_1 \cdot \textbf{r}} \end{pmatrix}.
  \label{blochform}
\end{equation}
We take $\vec k = 0$ to correspond to the K point in the moir\'{e} Brillouin zone (BZ).

There is guaranteed to be a zero-mode at the K point, $ \textbf{k} = 0$.  This is because there is one for $\alpha = 0$ and it is guaranteed to exist at all $\alpha$ because it is protected and pinned by symmetries \cite{Tarnopolsky2019}. Another way to understand this is by noting that the two different wavefunctions at K transform under conjugate representations of $C_3$ which are glued together by $C_2 \mathcal T$ symmetry \cite{Po2018}. The chiral symmetry ensures that this pair of levels remains at zero (since it interchanges positive and negative energy states). We therefore have $D(\textbf{r}) \psi_K( \textbf{r}) = 0$, where $\psi_K( \textbf{r})$ is the $K$ point wavefunction for positive chirality.

To find the other zero modes, we use that $D(\textbf{r})(F(z)\psi( \textbf{r})) = 0$ if $D( \textbf{r}) \psi( \textbf{r}) = 0$ for any holomorphic function $F(z)$ \cite{Tarnopolsky2019}.  Since the K-point wavefunction is a zero mode, we may generate other zero modes by multiplying it by some $F(z)$.  The wavefunctions $F(z) \psi_K(\textbf{r})$ are generically not normalizable since Liouville's theorem implies that a non-constant $F(z)$ cannot be bounded.  However, at the magic angles, we have $\psi_K( \textbf{r}_0) = 0$ (both components), for $ \textbf{r}_0 = ( \textbf{a}_1 - \textbf{a}_2)/3$\,\cite{Tarnopolsky2019}.  We can therefore let $F(z)$ be meromorphic, provided it only has poles when $z = z_0 = (a_1-a_2)/3$ and translates by lattice vectors $z \to z+a_1$, $z \to z+a_2$.  Here we use a non-bold letter to denote the complex version of a vector, e.g.:
\begin{equation}
  a_{1,2} = (\textbf{a}_{1,2})_x + i ( \textbf{a}_{1,2})_y. \\
  \label{cpxlattice}
\end{equation}
The chiral-MATBG wavefunctions are then of the form \begin{equation}
  \psi( \textbf{r}) = \frac{f(z)}{\th_1((z-z_0)/a_1|\omega)} \psi_K( \textbf{r}),
  \label{flatbandwfs}
\end{equation}
where $\omega = e^{i\phi} = a_2/a_1$.  Here $f(z)$ is a holomorphic function that is not and does not need to be bounded.  The function
\begin{equation}
\vartheta_1(u|\tau) = -i\sum_{n=-\infty}^\infty(-1)^n e^{\pi i\tau(n+\frac{1}{2})^2 + \pi i(2n+1)u}
\end{equation}
is the Jacobi theta function of the first kind, is odd in $z$, and has zeros when $z = m+n\omega$ where $m,n$ are integers \cite{Kharchev2015}.  Under translations it satisfies \cite{Kharchev2015}
\begin{equation}
    \begin{aligned}
      \vartheta_1(u+1|\tau) & = -\vartheta_1(u|\tau) \\
      \vartheta_1(u+\tau|\tau) & = - e^{-i\pi \tau -2\pi i u} \vartheta_1(u|\tau).
     \end{aligned}
     \label{thetatrans}
\end{equation}

For Bloch-periodic states we choose 
\begin{equation}
  \begin{aligned}
    f(z) & = e^{2\pi ik_1z/a_1}\th_1\left(\frac{z-z_0}{a_1} - \frac{k}{b_2}\bigg|\omega\right), \\ 
\end{aligned}
  \label{blochstates}
\end{equation}
where $k = k_x + ik_y$.
The Bloch-periodicity then follows from the behavior of the theta functions under translations \eqref{thetatrans}\cite{Kharchev2015}. The conventions here and that of Ref.~\onlinecite{Tarnopolsky2019} agree up to $\textbf{k}$-dependent normalizations. We choose this gauge because the moir\'{e}-periodic functions
\begin{equation}
  \begin{aligned}
  u_{ \textbf{k}}( \textbf{r}) & = e^{-i \textbf{k} \cdot \textbf{r}}\psi_{ \textbf{k}}( \textbf{r}) \\
  & = e^{-2\pi i r_2 k/b_2} \frac{\th_1\left(\frac{z-z_0}{a_1}-\frac{k}{b_2}|\omega\right)}{\th_1\left(\frac{z-z_0}{a_1}|\omega\right)} \psi_K(\textbf{r}).
\end{aligned}
  \label{hol_periodic}
\end{equation}
are holomorphic in $k$. The existence of such a gauge originates from the fact that $D(\vec r)$ only has antiholomorphic derivatives.  The zero mode equation $D(\vec r) \psi_{\vec k}(\vec r) = 0$ can then be recast as
\begin{equation}
    \begin{pmatrix} -k_\theta^{-1} 2i\bar \partial_k& \alpha U(\textbf{r})  \\ \alpha U(-\textbf{r}) & -k_\theta^{-1}2i\bar \partial_{k+q_1} \end{pmatrix} u_{\vec k}(\vec r) = 0,
\end{equation}
where $\bar \partial_k = \bar \partial + k$. The function $u_{\vec k}(\vec r)$ can then be chosen to only depend on $k$ and not $\bar k$, as it is the only zero mode for a matrix that only depends on $k$.

Introducing a sublattice potential that is equal on both layers then creates a promising situation for a fractional Chern insulator state without changing wavefunctions \eqref{flatbandwfs}.  Indeed, such a potential has the effect \begin{equation}
  \mathcal{H}_{\rm chiral} \to \mathcal{H}_{\rm chiral} - \delta \sigma_z,
  \label{sublattice}
\end{equation} where $\delta = \Delta/E_0 = \alpha \Delta/w$ is the dimensionless potential strength.  Because the flat band states can be chosen to be eigenstates of $\sigma_z$, introducing this potential splits the two bands without changing the wavefunctions.  A potential on a single layer would change the wavefunctions, but it would still not introduce mixing between the flat bands because it commutes with $\sigma_z$.  It would therefore only change the wavefunctions via mixings with higher bands.  Such effects are suppressed by factors of $\delta/E_0$.  For the rest of this paper, we therefore work with split flat bands and focus on the lower band with wavefunctions \eqref{flatbandwfs}, \eqref{blochstates}.  

The positive and negative chirality bands have Chern number $-1$ and $+1$ respectively.  As they are perfectly flat and separated in the presence of a sublattice potential \eqref{sublattice}, we attain ideal band energetics for hosting a FCI state by fractionally filling the positive chirality band.  We assume here that the system is spin and valley polarized for simplicity, which in realistic systems is achieved through interaction-induced flavor polarization \cite{XieMacDonald, Liu19, Bultinck19, Zhang19, DavidGG, AndreaYoung}, though spin unpolarized states can be included in the same way as in the LLL.  In the next section we discuss the near-ideal band geometry of the flat bands.

\section{Band Geometry of the flat bands}\label{bandgeometry}

In this section we discuss the band geometry of the wavefunctions \eqref{flatbandwfs}.  In particular, we will show that chiral-MATBG has near-ideal band geometry due to holomorphicity in momentum space \eqref{hol_periodic}.  We emphasize that band geometry is a strong predictor of FCI stability: it appears in many analytical arguments for ideal descriptions and its deviation from ideality strongly correlates with the many body gap \cite{Jackson2015}.

Before we move forward, we review the various band-geometric conditions.  All geometric information is encapsulated in the complex tensor
\begin{equation}
  \eta_{ab}(\vec k) = \frac{\bra{\partial_{k_a} u_{\vec k}}\ket{\partial_{k_b} u_{\vec k}}}{\norm{u_k}^2} - \frac{\bra{\partial_{k_a} u_{\vec k}}\ket{u_{\vec k}}\bra{u_{\vec k}}\ket{\partial_{k_b} u_{\vec k}}}{\norm{u_k}^4}
  \label{etatensor}
\end{equation}
which does not depend on the $\vec k$-dependent norm or phase of the wavefunctions $u_\vec{k}$.  We use moir\'{e}-periodic wavefunctions in these overlaps as is standard.  The imaginary and real components of this tensor respectively capture the phase and magnitude of Bloch overlaps under varying $\vec k$ and correspond to the Berry curvature $\Omega(\vec k)$ and the Fubini-Study metric $g_{ab}(\vec k)$ respectively: 
\begin{equation}
\begin{aligned}
  \Omega(\vec k) & = -2\Im \eta_{xy}(\vec k) = 2 \Im \eta_{yx}(\vec k), \\
  g_{ab}(\vec k) & = \Re \eta_{ab}(\vec k).
\end{aligned}
  \label{fubinistudy}
\end{equation}

To motivate these geometrical considerations, we review the GMP algebra in the LLL \cite{GMP}.  The many-body quantum Hall Hamiltonian projected to the LLL can be written as 
\begin{equation}
    H_{\rm LLL} = \sum_{q} V(q) e^{-q^2\ell_{B}^2/2}\bar \rho_{\vec q} \bar \rho_{-\vec q},
\label{LLLdensity}
\end{equation}
where $\ov \rho_{\vec q}$ is obtained from the density operator $\rho_{\vec q} = \sum_i e^{i\vec q\cdot \hat{\vec r}_i}$by LLL projection: $\mathcal P \rho_q \mathcal P =e^{-q^2\ell_{B}^2/4}\sum_ie^{i\vec{q} \cdot \hat{\vec R}_i} =   e^{-q^2\ell_{B}^2/4} \ov \rho_{\vec q}$.  The coordinate $\hat{\vec R}_i$ 
is the guiding center coordinate for the $i$'th particle and the pre-factor of 
$e^{-q^2\ell_{B}^2/4}$ is known as the LLL form factor. Because the guiding center coordinates do not commute, 
$\left[\hat R_x,\hat R_y\right] = -i \ell_B^2$,
the projected densities do not either and can be shown to obey the GMP algebra \cite{GMP}
\begin{equation}
    \left[\ov \rho_{\vec q},\,\ov{\rho}_{\vec q '}\right] = 2 i \sin \left(\frac{\vec q\times \vec q' }{2}\ell_{B}^2\right) \ov \rho_{\vec q+\vec q'}.
\label{FullGMP}
\end{equation}

It was noticed early on that to mimic Landau levels, Berry curvature should be flat.  In the context of the GMP algebra, this allows one to reproduce the algebra at small $\vec q$, in particular one obtains
\begin{equation}
    \left[\tilde \rho_{\vec q},\,\tilde{\rho}_{\vec q '}\right] = i \Omega \vec q\times \vec q' \tilde \rho_{\vec q+\vec q'} + O(q^3),
\label{lowestorderGMP}
\end{equation}
where $\tilde \rho_{\vec q} = \sum_\bk c_\bk^\dagger \lambda_\bq(\bk) c_{\bk + \bq}$, with $\lambda_\bq(\bk) = \langle u_\bk|u_{\bk+\bq}\rangle$ \cite{Zhang2018}, are the projected densities in the Chern band such that $H_{\rm Chern} = \sum_q V(\vec q) \tilde \rho_{\vec q}\tilde \rho_{-\vec q}$
\cite{Parameswaran2012}. The algebra \eqref{lowestorderGMP} matches \eqref{FullGMP} in the small $q$ limit, with $\Omega$ as the analogue of $\ell_B^2$. The algebra no longer closes, however.

One must add conditions on the metric $g(\vec k)$ to reproduce the algebra in full. The full algebra \eqref{FullGMP} is recovered, with $\Omega$ taking the role of $\ell_B^2$, if and only if the metric is $\vec k$-independent and satisfies 
\begin{equation}
  \det \eta(\vec k)  = 0 \iff 4 \det g(\vec k) = \abs{\Omega(\vec k)}^2,
  \label{detcond}
\end{equation}
for each $\vec k$ in the BZ.  In particular, it is recovered for the operators $\ov \rho_{\vec q} = e^{q^2\Omega/4} \tilde \rho_{\vec q}$, such that LLL physics is reproduced \cite{Roy2014,Parameswaran2013}.  Other Landau levels crucially have a different form factor.

The condition \eqref{detcond} holds for all two band tight binding models, although the metric is usually {\em not} constant and its eigenvectors typically {\em  vary} between different values of $\vec k$.  The condition \eqref{detcond}  is referred to as the determinant condition or the ideal droplet condition. If the metric varies only in magnitude then one may diagonalize it, which leads to a stronger version of \eqref{detcond},
\begin{equation}
  g_{ab}(\vec k) = \frac{1}{2}\abs{\Omega(\vec k)}\delta_{ab}.
  \label{trcond}
\end{equation}
This condition is considerably harder to satisfy and is referred to as the trace condition, as it saturates the inequality $\tr g \geq \abs{\Omega(\vec k)}$ \cite{Roy2014}. It is also known as the isotropic ideal droplet condition.

The real-space periodic Bloch states \eqref{blochstates} have the remarkable property that they are holomorphic in $k_x + ik_y$.  It was noticed in Ref.~ \onlinecite{Claassen2015} that this type of holomorphicity implies \eqref{trcond} in tight-binding models.  The proof is similar in our continuum model, which we show now for completeness.  Holomorphicity in $k_x + ik_y$ implies $\partial_{k_x} u_{\vec k} = (\partial + \bar \partial) u_{\vec k} = \partial u_{\vec k}$, where the holomorphic derivatives are now in $k$-space, $2\partial = \partial_{k_x} -i\partial_{k_y}$.  Similarly, $\partial_{k_y} u_{\vec k} = i \partial u_{\vec k}$.  We therefore have 
\begin{equation}
  \eta(\vec k) \propto \begin{pmatrix} 1 & i \\ -i & 1 \end{pmatrix},
  \label{holomorphic_eta}
\end{equation}
from which \eqref{trcond} follows.

Refs.~\onlinecite{Claassen2015,Lee2017} previously considered ideal models satisfying \eqref{trcond} by writing down wavefunctions that are holomorphic in $k_x + ik_y$. If one imposes periodic boundary conditions in $\vec k$-space, these are  elliptic functions which have $\abs{C} \geq 2$ because they have two or more poles in the BZ \cite{Jian2013}.  The authors of Refs.~\onlinecite{Claassen2015,Lee2017} thus restricted their attention to elliptic function models with $\abs{C} \geq 2$. However, periodic boundary conditions in $\vec k$-space are not required for a well defined Berry curvature or Chern number. The chiral-MATBG Bloch states \eqref{blochstates} are not periodic, and they were therefore missed in the above studies of ideal models satisfying \eqref{trcond}. 

\begin{figure}
    \centering
    \includegraphics[width = 0.5\textwidth]{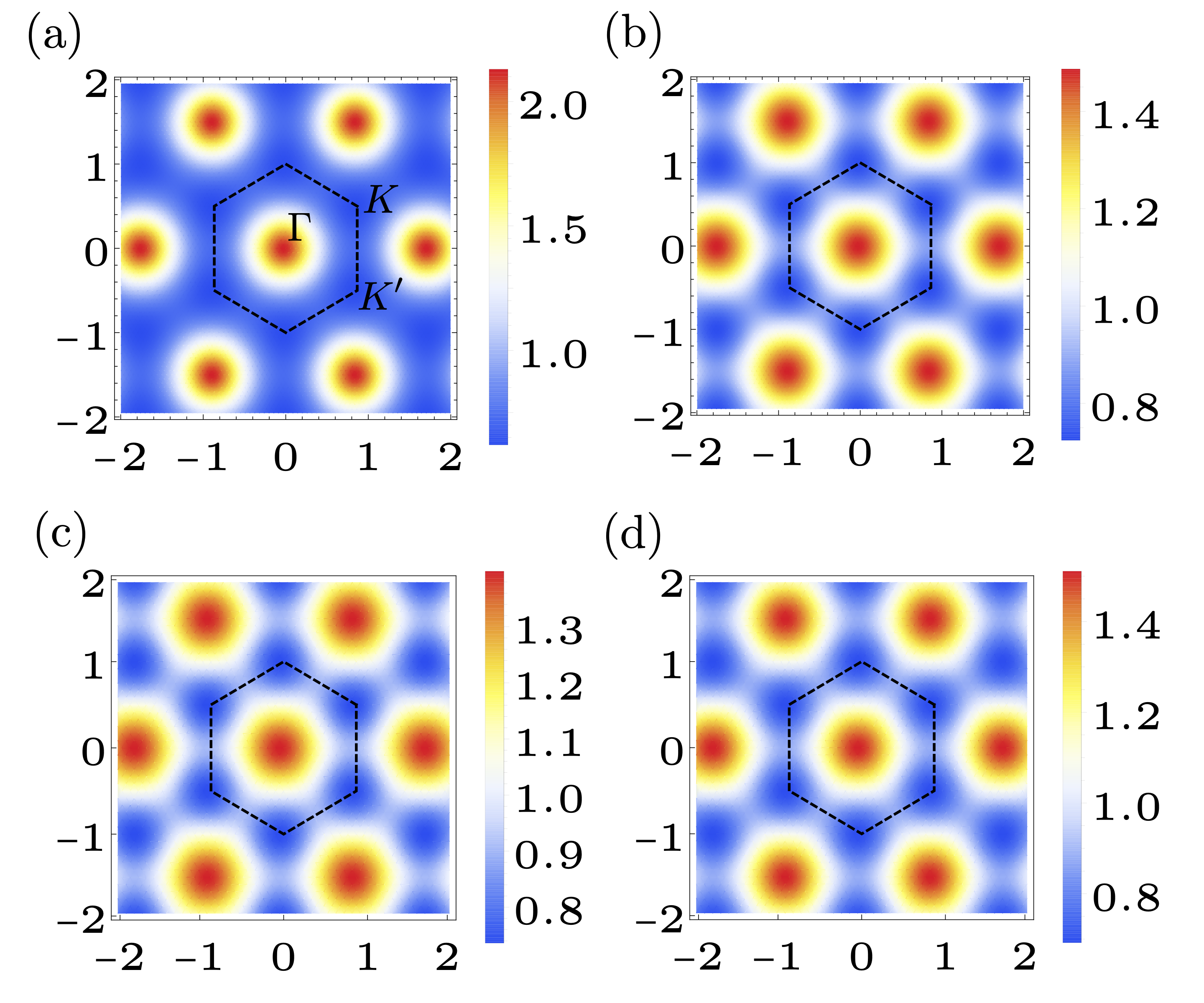}
    \caption{Berry curvature $\abs{\Omega(\bk)}$ as a function of momentum in chiral-MATBG for the first four magic angles.  Panels (a)-(d) correspond to $\alpha = 0.585,\, 2.221,\, 3.751,\, 5.276$ respectively.
}
    \label{Falpha}
\end{figure}

The only non-ideal aspect of chiral-MATBG is inhomogeneous Berry curvature.  Because Berry curvature spreads out in the chiral limit (and outside of the chiral limit in the presence of a sublattice potential), this is not a large effect.   We can quantify Berry curvature fluctuations through 
\begin{equation}
  F = \sqrt{\int_{\rm BZ} \frac{d^2 \vec k}{A_{\rm BZ}} \left( \frac{\Omega}{2\pi} - C \right)^2},
  \label{berryfluc}
\end{equation}
where $A_{\rm BZ}$ is the area of the BZ and $C = -1$ is the Chern number for the positive chirality band in this valley.  For the first through fourth magic angles we have $F = 0.373, \, 0.197, \, 0.173, \, 0.213$ respectively. This is illustrated in Fig.~\ref{Falpha} which shows the distribution of Berry curvature in momentum space for the first four magic angles showing relatively uniform Berry curvature, particularly for the second and third magic angles.

To compare with other models it will be useful to quantify violation of \eqref{trcond}. We use the BZ average $\ov T = \langle T(\vec k) \rangle_{\rm BZ}$ of the (non-negative \cite{Roy2014}) quantity 
  \begin{equation}
    T(\vec k) = \tr g(\vec k) - \abs{\Omega(\vec k)},
    \label{trcondviolate}
  \end{equation}
  as in Ref.~\onlinecite{Jackson2015}.  If \eqref{trcond} is satisfied, $\ov T = T(\vec k)$ = 0.
  
  Because chiral-MATBG exactly satisfies \eqref{trcond} and has small $F$, it is one of the most promising models to host FCI states, even compared to fine-tuned models. The study of ideal FCI models has shown that the confluence of flat bands, constant Berry curvature, and the ideal metric \eqref{trcond} are very rare.  These geometric qualities are important: Ref.~\onlinecite{Jackson2015} finds that the many body gap for FCI states is strongly negatively correlated with $F$ and $\ov T$.   Because chiral-MATBG has small $F$ and exactly satisfies \eqref{trcond}, it is one of the most promising systems for FCI states.  To argue this, we compare with models in Ref.~\onlinecite{Jackson2015}.  
The results are shown in Table \ref{geomtable}; chiral-MATBG is just as promising as fine-tuned lattice models with only nearest neighbor hoppings, and next nearest neighbor hoppings for the Haldane model, included \cite{Jackson2015}.  If arbitrary length hoppings with $>3$ bands are chosen then both $F$ and $\overline{T}$ can be made arbitrarily small \cite{Lee2017}.
It is remarkable that chiral-MATBG compares well to fine-tuned models without any parameter optimization.
\begin{table}[htb]
  \begin{ruledtabular}
    \begin{tabular}{ccccc}
    & \multicolumn{4}{c}{Chiral-MATBG} \\

       & $\alpha_1$ & $\alpha_2$ & $\alpha_3$ & $\alpha_4$ \\
      \hline
      \hline
      $F$ & $0.373$  & $0.197$ & $0.173$ & $0.213$   \\

      $\overline T$ & $0$  & $0$ & $0$ & $0$ \\
      \hline
      \hline
          & \multicolumn{4}{c}{F-minimizing Tight binding models\cite{Jackson2015}} \\

       & Haldane & Haldane + NNN & Kagome & Ruby \\
      \hline
      \hline
      $F$           & $0.993$  & $0.147$ & $0.222$ & $0.134$ \\

      $\overline T$ & $0.467$  & $3.44$ & $0.796$ & $0.172$ \\
      
    \end{tabular}
  \end{ruledtabular}
  \caption{Deviations from ideal band geometry $F$ and $\overline{T}$, see \eqref{berryfluc} and \eqref{trcondviolate}.  Chiral-MATBG compares well to tight binding models with nearest neighbor hoppings and the Haldane model with next nearest neighbor (NNN) hoppings.  These tight binding models have been optimized to minimize $F$\cite{Jackson2015}.}
    \label{geomtable}
\end{table}

While the above analysis suggests that chiral-MATBG is promising for FCI ground states, to make analytical progress that shows chiral-MATBG can host an FCI ground state one must pursue pseudopotential-like arguments.  For the FQHE, Haldane \cite{Haldane1983} used rotation and magnetic translation symmetry to reduce all interaction matrix elements to a discrete collection of numbers $V_m$, where $m$ is the (conserved) relative angular momentum of a pair of particles.  Breaking these symmetries splits the degeneracy between these matrix elements\cite{Bergholtz2013}.  One expects that the FQHE can be recovered, say at $\nu = 1/3$, if the gap between $V_1$ and $V_3$ is much larger than the splitting of each.  In the next section we pursue an argument similar to this: we construct a Laughlin state on a real-space torus geometry that is a zero-energy ground state of pseudopotential-like interactions.  These pseudopotential-like interactions will not be described by a set of discrete matrix elements and will be split into bands \cite{Bergholtz2013}.  For a screened Coulomb interaction with screening length much less than $a$, we will argue that these bands are well separated.

\section{Laughlin state on a Moir\'{e} Torus} \label{Torus}

In this section we investigate the chiral model on a moir\'{e} torus and derive a Laughlin-like wavefunction.  We closely follow Haldane and Rezayi's derivation for the FQHE \cite{Haldane1985}.  In fact, we obtain an exact mapping onto their derivation which can be anticipated as follows.  Previous studies of FCIs have identified the dictionary $\ell_B^2 \to \langle \Omega \rangle$ for translating from the FQHE to FCIs,  where $\ell_B^2$ is the squared magnetic length and $\langle \Omega \rangle$ is the average Berry curvature (for example: Refs.~\onlinecite{Parameswaran2012,Parameswaran2013,Claassen2015}).  One may understand this in the context of Harper-Hofstadter models, where both a magnetic length and average Berry curvature is present, such that $2\pi\ell_B^2$ is equal to the area of an effective unit cell where translation operators commute.  

Applying this to TBG, we replace $2\pi \ell_B^2$ with the area of the moir\'{e} cell: \begin{equation}
  2\pi \ell_B^2 \to \norm{\vec a_1 \times \vec a_2} = \frac{\sqrt{3}}{2}a^2
  \label{lengthscales}
\end{equation}
where $a = \norm{\vec a_{1,2}}$ is the moir\'{e} lattice constant. We now apply this dictionary to obtain TBG on a torus by suitably modifying each step of Ref. \cite{Haldane1985}.  

The parallelogram that defines our torus will be spanned by the vectors $\vec L_1 = N_1 \vec a_1$ and $\vec L_2 = N_2 \vec a_2$, where $N_1$ and $N_2$ are positive integers so that there is an integer number of unit cells in each direction.  This is the equivalent of magnetic flux quantization through a torus; indeed under the replacement \eqref{lengthscales} the flux quantization $\abs{\vec L_1 \times \vec L_2} = 2\pi\ell_B^2 N_s$, where $N_s$ is an integer, becomes $N_1N_2=N_s$.

We specify generic twisted boundary conditions \begin{equation}
\begin{aligned}
  \psi(\vec r + N_1\vec a_1)  & =U^{N_1}e^{i \phi_1 + i\pi N_1}\psi(\vec r), \\
  \psi(\vec r + N_2\vec a_2)  & =U^{N_2}e^{i \phi_2 + i\pi N_2}\psi(\vec r),
  \end{aligned}
  \label{twistedbcs}
\end{equation}
where $U = \rm{diag}(1,\bar \omega)$ accounts for the extra phase in the second layer \eqref{blochform}. The factors $e^{i\pi N_{1,2}}$ will conveniently absorb the signs $(-1)^{N_{1,2}}$ in \eqref{thetatrans} from $z \to z+ N_{1,2}a_{1,2}$ in the upcoming discussion. Both extra factors are unimportant and disappear when $N_{1,2}$ are multiples of $6$.

For the wavefunctions \eqref{flatbandwfs}, the twisted boundary conditions imply
\begin{equation}
  \begin{aligned}
    f(z+L_1) & = e^{i\phi_1} f(z), \\
      f(z+L_2) & = e^{i\phi_2}e^{-2\pi iN_2^2\omega - 2\pi iN_2\frac{z-z_0}{a_1}}f(z),
  \end{aligned}
  \label{holbcs}
\end{equation}
where we used the behavior of $\th_1((z-z_0)/a_1|\omega)$ under translations \eqref{thetatrans}\cite{Kharchev2015}. These conditions \emph{exactly} match those in Ref.~\onlinecite{Haldane1985} provided we identify $N_s = N_1 N_2$ and $\tau = N_2 \omega/N_1$. Our $z$ coordinate is a dimensionful and shifted version of Haldane's. The dictionary between the problems arises because our $1/\vartheta_1$ factor has similar properties under translations to the factor $e^{-y^2/2\ell_{B}^2}$ that appears in QHE in the Landau gauge $\vec A = -By\hat{\vec{x}}$, a fact that is exploited further and made more precise in the next section. Indeed, everything from here onward is the same as in Ref.~\onlinecite{Haldane1985} if we use these identifications.  In particular, we write a Jastrow ansatz for the many-body wavefunction of $N_e$ electrons \begin{equation}
  \begin{aligned}
    \Psi(\{\vec r_i\}) & = F(\{z_i\}) \prod_{i=1}^{N_e} \frac{\psi_K(\vec r_i)}{\th_1\left(\frac{z_i-z_0}{a_1}|\omega\right)}, \\ 
    F(\{z_i\}) & = G(Z)\prod_{i<j} g(z_i - z_j),
  \end{aligned}
  \label{manybody}
\end{equation}
where $Z = \sum_{i} z_i$. The boundary conditions \eqref{holbcs} imply \cite{Haldane1985}\begin{equation}
  \begin{aligned}
    g(z+L_1) & = \eta_1g(z), \\
    g(z+L_2) & = \eta_2 e^{-2\pi i(N_s/N_e)z/L_1}g(z),
\end{aligned}
  \label{laughlinbcs}
\end{equation}
for some $z$-independent constants $\eta_{1,2}$.  

The integral of $\frac{d}{dz}\ln(g(z))$ around the parallelogram defining the moir\'{e} torus is $2\pi i$ times the number of zeros of $g(z)$.  The boundary conditions above then imply that the number of zeros is $m = N_s/N_e$, which must then be an integer.  If we demand that all the zeros are at the same point, we obtain the Laughlin factor \begin{equation}
  g(z) = \left( \th_1(z/L_1|\tau) \right)^m.
  \label{laughlin}
\end{equation}
We must have $m$ odd for fermion antisymmetry to hold. 

The function \eqref{laughlin} vanishes as $z^m$ when two particles approach each other.  
This ensures that this wavefunction will be a zero energy eigenstate of any ``pseudopotential'' 
\begin{equation}
  V(\vec r) = \sum_{n=0}^{m'} v_n (a\nabla)^{2n} \tilde{\delta}(\vec r),
  \label{PseudoTorus}
\end{equation}
where $m' < m$ and 
\begin{equation}
\tilde{\delta}(\vec r) = \sum_{\alpha, \beta \in \mathbb{Z}} \delta(\vec r - \alpha \vec L_1 - \beta \vec L_2 )
  \label{torusdelta}
\end{equation}
is a delta function that respects the periodicity of the torus.  We include the factors of $a$ with $\nabla$ since all band-projected states vary on the scale of $a$. Note that only the odd-$n$ terms matter for fermions.

In particular, because all band-projected states are smeared on the scale of the moir\'{e} length, a Coulomb interaction screened at distances much shorter than the moir\'{e} length will be well approximated as $V(\vec r) \approx  v_0 \delta(\vec r) + v_1 a^2\nabla^2 \delta(\vec r)$, and \eqref{laughlin} suffers no energy cost from this potential for $m = 3$ (though we cannot identify the coefficients with matrix elements of relative angular momentum eigenstates).  Indeed, an expansion of a screened Coulomb interaction with screening length $a_C$ generically leads to $v_n \sim (a_C/a)^{2n} v_0$, with $v_0 \sim e^2a_C/\varepsilon$.  Even with $a_C = a/2$ we have $v_1/v_3 \sim 16$, leading to a strong preference for a state that is a zero mode of $v_1 a^2 \nabla^2\tilde{\delta}(\vec r)$ like \eqref{laughlin} is for $m = 3$. 

We note that angular momentum conservation or magnetic translation symmetry are not necessary for these arguments; they simply make the discussion on a plane or sphere more streamlined as symmetry arguments alone dictate that one only needs to care about Haldane pseudopotentials \cite{Haldane1983}. Matrix elements here in contrast will have dependence on the center of mass coordinate (due to lack of magnetic translation symmetry), but this should not matter as for sufficiently short range potentials this dependence will be dwarfed by the gap between $v_1$ and $v_3$.  

As in Ref.~\onlinecite{Haldane1985}, one may proceed to derive boundary conditions for the center of mass piece of the wavefunction 
\begin{equation}
  \begin{aligned}
    \frac{G(Z+L_1)}{G(Z)} & = e^{i \phi_1 + i\pi(N_s-m)}, \\
    \frac{G(Z+L_2)}{G(Z)}&  =  e^{-\pi i m \tau - 2\pi i m \frac{Z-Z_0}{L_1} + i\phi_2+i\pi(N_s-m)},
  \end{aligned}
  \label{com_bcs}
\end{equation}
where $Z_0 = N_e z_0$. The general solution to these boundary conditions is \begin{equation}
  G(Z) = e^{iKZ}\prod_{\nu = 1}^m \th_1((Z-Z_\nu)/L_1 | \tau),
  \label{comsoln}
\end{equation}
subject to the constraints 
\begin{equation}
  \begin{aligned}
    e^{iKL_1} & = (-1)^{N_s} e^{i\phi_1}, \\ 
    e^{2\pi i \sum_\nu (Z_\nu-Z_0)/L_1} & = (-1)^{N_s} e^{i\phi_2 - iKL_1\tau}.
\end{aligned}
  \label{comsoln2}
\end{equation}
From these expressions, one finds that there are $m$ linearly independent solutions which characterizes the $m$-fold degeneracy of Laughlin-like FCI states on the torus.  Furthermore, these solutions braid into each other upon adiabatically inserting a flux quantum through slowly increasing $\phi_1$ or $\phi_2$.  One may go further and discuss quasihole excitations as well, as in Ref.~\onlinecite{Haldane1985}.

\section{Flat TBG band as a generalized LLL}
\label{curvedLLL}

In this section, we interpret the previous results by showing that the chiral-MATBG wavefunctions \eqref{flatbandwfs} also arise in a generalized LLL where the magnetic field is inhomogeneous.  To do this we note that only the magnitude of $\psi_K(\vec r)/\vartheta_1((z-z_0)/a_1|\omega)$ enters in band-projected matrix elements of density-density operators: the position dependent layer polarization and phase has no effect.  We will find generalized LLL wavefunctions that match the chiral-MATBG wavefunctions \eqref{flatbandwfs}, except with $\psi_K(\vec r)/\vartheta_1((z-z_0)/a_1|\omega)$ replaced by its magnitude.

Generally an inhomogeneous magnetic field leads to Landau level dispersion and wavefunctions that are not holomorphic.  There are two situations that we found where this is not the case and states in the LLL can be defined as zero modes of the antiholomorphic canonical momentum $\ov \Pi = (\Pi_x + i \Pi_y)/2$ where $\vec \Pi = \vec p - e \vec A$ and $\nabla \times \vec A = B(\vec r)$.  

One situation is that of a Dirac particle with the Hamiltonian \begin{equation}
  H = \vec \sigma \cdot \vec \Pi = 2 \begin{pmatrix} 0 & \Pi \\ \ov \Pi & 0 \end{pmatrix}.
  \label{diracQH}
\end{equation}
The zeroth Landau level is then sublattice polarized and a zero mode of $\ov \Pi$.  

The other situation is that of a nonrelativistic particle in a K\"{a}hler geometry.  Here, one has both a curved space and an inhomogeneous magnetic field, such that the magnetic field is proportional to the K\"{a}hler form \cite{Can2014,Douglas2009,Klevtsov2014,Gromov2017}.  
However, this case comes with the additional stipulation that $B$ cannot vary in sign (else $\sqrt{\det g}$ would), and so the Dirac model is more general.  We will see that the inhomogeneous $B(\vec r)$ of interest to us does vary in sign within the moir\'{e} cell.

We now need to discuss the zero modes of $\ov \Pi = -i\bar\partial - e\bar A$.  We take $\bar{A} = iB_0 \bar \partial K/4$ 
for some 
$K(\vec r)$.  The magnetic field is then $B(\vec r) = B_0\partial \bar \partial K$.  The states in the generalized LLL are 
\begin{equation}
  \psi_{\rm LLL}(\vec r) = f(z)e^{-K(\vec r)/4\ell_{B_0}^2},
  \label{inhomLLLstates}
\end{equation}
where $\ell_{B_0} = \sqrt{\hbar/eB_0}$.  

\begin{figure}[htb]
  \centering
  \includegraphics[width=0.5\textwidth]{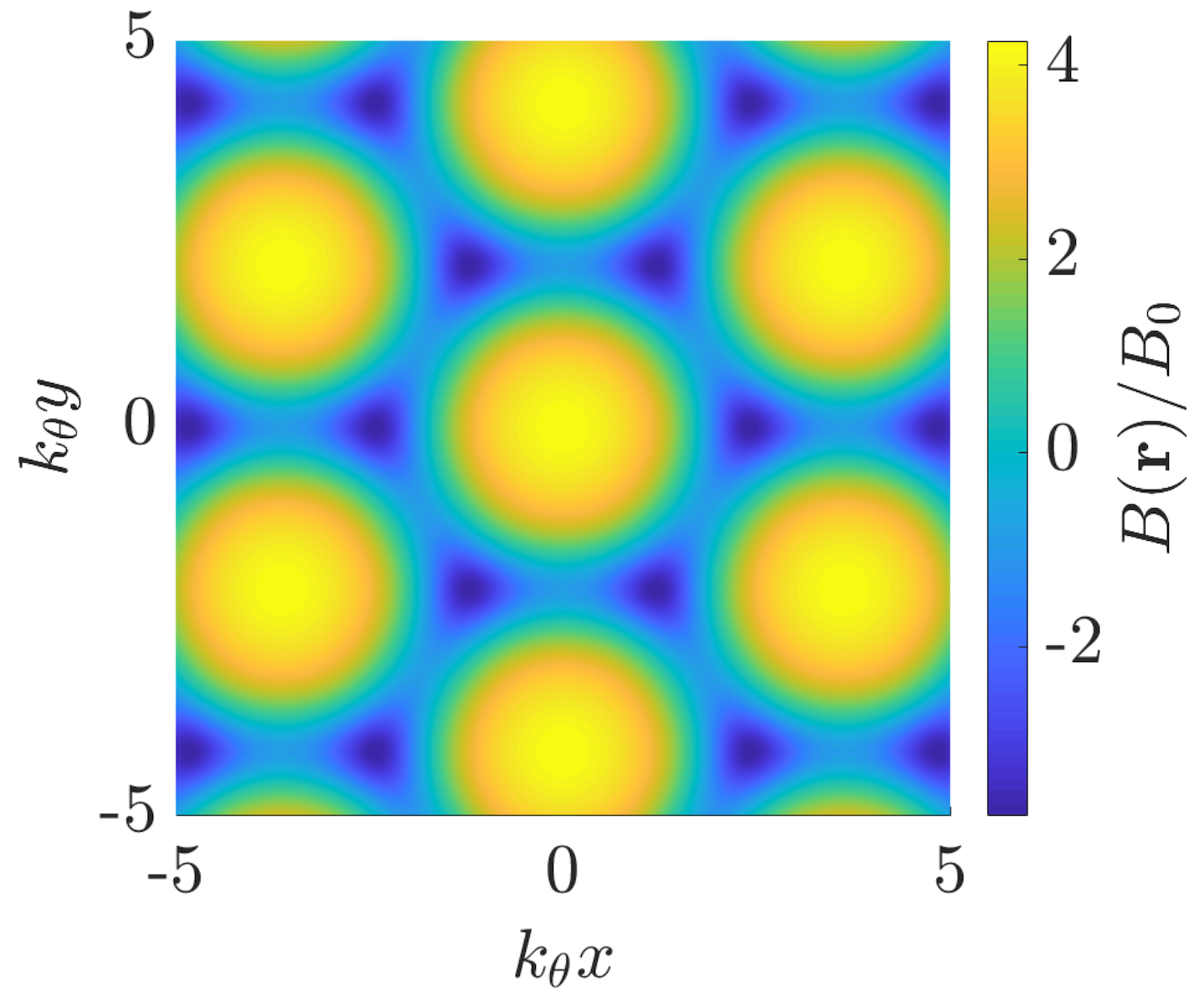}
  \includegraphics[width=0.5\textwidth]{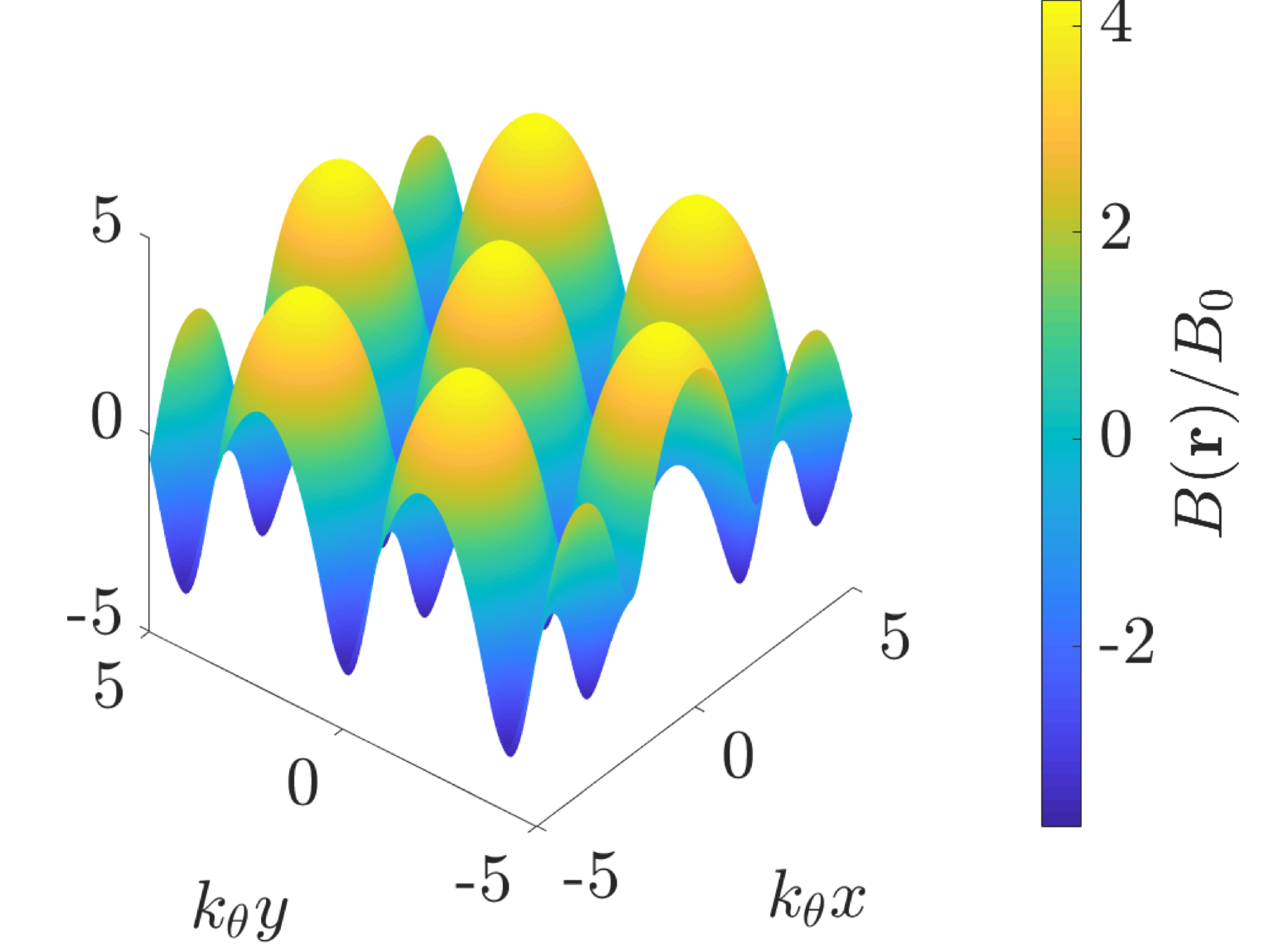}
  \caption{Effective magnetic field $B(\vec r)$ such that TBG states are LLL states.  The magnetic field has moir\'{e}-periodicity stemming from its relation to TBG wavefunctions \eqref{inhomB}}.
  \label{fig:magfield}
\end{figure}

We will match \eqref{inhomLLLstates} with \eqref{flatbandwfs} for a suitable $K$ that we now define. 
We set $2\pi \ell_{B_0}^2 = a^2\sqrt{3}/2$ so that there is one flux quantum per moir\'{e} cell, as in \eqref{lengthscales}.
We also decompose $K = K_0 + K_\lambda$ so that 
\begin{equation}
  K_0 = \frac{3}{2}a^2(r_2+1/3)^2 
  \label{constantgauge}
\end{equation}
satisfies $\partial \bar \partial K_0 = 1$ and $K_\lambda$ generates mean-zero spacial variation $B(\vec r)-B_0 = B_0\partial \bar \partial K_\lambda$.  The numerical shift in the Landau gauge for $K_0$ was chosen to match the quasiperiodicity of the magnitude of the wavefunctions \eqref{flatbandwfs} under translations $z \to z+a_2$.

For later convenience we define 
\begin{equation}
  \begin{aligned}
  \abs{\psi_0}^2 & = e^{-K_0/2\ell_{B_0}^2}\abs{\vartheta_1\left(\frac{z-z_0}{a_1}\bigg|\omega\right)}^2,  \\ 
  & = e^{-\pi\sqrt{3}(r_2 +1/3)^2}\abs{\vartheta\left(\frac{z-z_0}{a_1}\bigg|\omega\right)}^2,
\end{aligned}
  \label{idealK}
\end{equation}
The function $\abs{\psi_0}^2$ is positive and moir\'{e}-periodic.  We will see that $\abs{\psi_0}^2$ represents an ``ideal'' version of $\abs{\psi_K}^2$. 

The $K_\lambda$ term generates the intra-unit-cell variations of $B$ and breaks continuous magnetic translation symmetry down to translations by moir\'{e} vectors.   It is given by 
\begin{equation}
  \begin{aligned}
    K_\lambda & = 2\ell_{B_0}^2 \log \abs{\frac{\psi_0}{\psi_K}}^2 \\
    & =  \frac{\sqrt{3}}{2\pi}a^2 \log \abs{\frac{\psi_0}{\psi_K}}^2 
\end{aligned}
  \label{Klambda}
\end{equation}
such that the inhomogeneities are driven by $\psi_K \neq \psi_0$.  Substituting \eqref{constantgauge}, \eqref{idealK}, and \eqref{Klambda} into \eqref{inhomLLLstates} one obtains the TBG wavefunctions \eqref{flatbandwfs} up to the previously discussed difference in phase and layer polarization.  

The inhomogeneous magnetic field is
\begin{equation}
  \frac{B(\vec r)}{B_0} =  1+ \frac{\sqrt{3}}{2\pi}a^2\partial\bar \partial\log \abs{\frac{\psi_0}{\psi_K}}^2,
  \label{inhomB}
\end{equation}
It is plotted in Fig. \ref{fig:magfield}. We note that while $B(\vec r)$ is strongly inhomogeneous and changes sign within a moir\'{e} cell, much of the variation is averaged over in matrix elements as all band projected states are delocalized over areas $\sim a^2$.

These results explain the success of the arguments in the previous section, where we constructed a Laughlin state on a torus using Haldane's arguments in Ref.~\onlinecite{Haldane1985}.  Indeed, one would expect this for any generalized LLL where the magnetic field is periodic, as the only difference in the wavefunctions from the ordinary LLL is a periodic factor $e^{-K_\lambda/2\ell_{B_0}^2}$.  This extra periodic factor does not change anything about the boundary conditions on the torus and so all arguments go through unaffected. 

We may relate the above approach to the momentum space geometry described in section \ref{bandgeometry} by carrying out this process in reverse.  I.e., one may consider any periodic magnetic field $B(\vec r)$ and the corresponding LLL wavefunctions \eqref{inhomLLLstates} with $K_0$ given by \eqref{constantgauge} and $\partial\bar \partial K_\lambda = (B(\vec r)-B_0)/B_0$ so that $\abs{\psi_K}^2 = \abs{\psi_0}^2 \exp(-K_\lambda/2\ell_{B_0}^2)$.  One may then use Bloch periodic states similar to \eqref{blochstates}, with the overall position dependent phase and layer polarization removed. Because the periodic states are holomorphic in $k_x + ik_y$ the condition \eqref{trcond} is exactly satisfied.  The Berry curvature will in general be inhomogeneous, as it is for chiral-MATBG.  

It is interesting to ask whether every FCI model that satisfies \eqref{trcond} can be described this way.  Indeed, there has been a lot of work on curved-space Landau levels including both first-quantized real-space Laughlin approaches and effective field theories, for example Refs.~\onlinecite{Can2014, Klevtsov2014,Douglas2009, Gromov2017}.  Despite their success in the FQHE, these techniques so far have seen little use in the FCI literature due to the discreteness of the lattice.  We argue that our work may change this state of affairs for models satisfying \eqref{trcond}; other models may be approached from these special models through adiabatic continuity.

\section{Deviations from Chiral model: $\kappa > 0$}
\label{realistic}

In this section we explore deviations from the chiral model: $\kappa > 0$. The parameter $\kappa$ in \eqref{tunneling} describes the relative strength between tunneling at AA sites and AB/BA sites, such that if $\kappa = 0$ AA tunneling is absent.  Lattice relaxation effects reduce AA tunneling and increase AB/BA tunneling so that $\kappa$ is smaller than the value of $1$ which was assumed in the original Bistritzer-Macdonald model \cite{Bistritzer2011}.  
Refs.~\onlinecite{Koshino2018,Carr2019} estimate $\kappa \approx 0.7-0.8$.   While the approximation $\kappa = 0$ may seem drastic, the sublattice-polarized flatband wavefunctions for the realistic model are adiabatically connected to those of the chiral model \cite{BultinckKhalaf}. Combined with the fact that the topological order of FCI states is stable to deformations up to some critical value, this suggests that an FCI state at $\kappa = 0$ may be adiabatically connected to potential FCI states at realistic values of $\kappa$.

\begin{figure*}[htb]
    \centering
    \includegraphics[width = \textwidth]{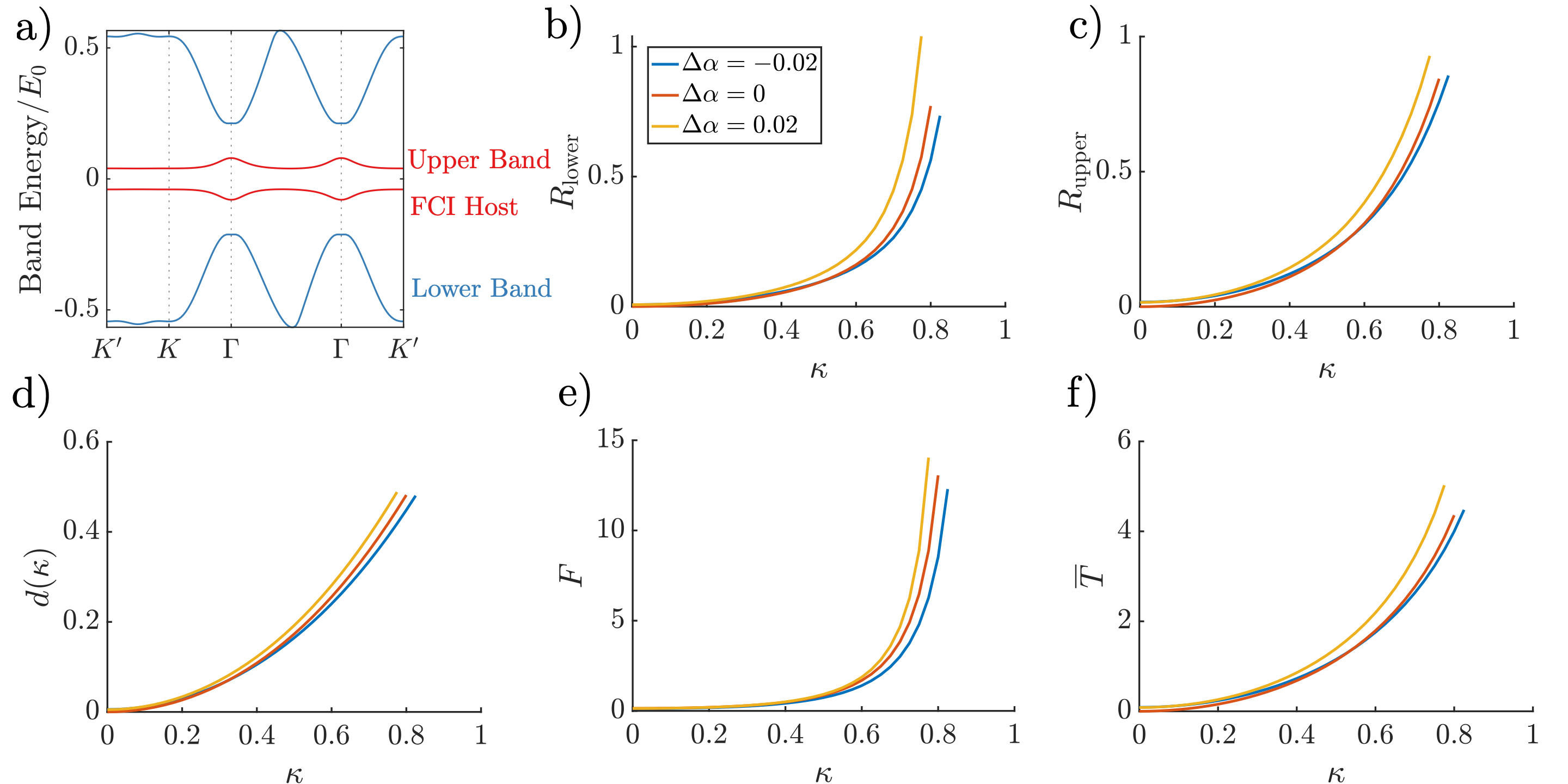}
    \caption{Band energetic and geometric quantities for $\kappa > 0$.  In (a) the four energy bands in a single valley that are closest to charge neutrality are shown at the first magic angle for $\kappa = 0.7$, $\delta = 0.1$, and $\Delta \alpha =0$.  The bottom red band is assumed to be fractionally filled and hosts the candidate FCI state.  The top red band and the bottom blue band are the upper and lower neighboring bands respectively.  The lower and upper inverse flatness ratios $R_{\rm lower}, \, R_{\rm upper}$, wavefunction distance $d(\kappa)$, Berry curvature inhomogeneity $F$, and average violation of \eqref{trcond} $\ov T$ are plotted in (b-f) respectively.  All become less ideal as $\kappa$ increases, though less so for $\Delta \alpha < 0$. Points for which the lower band gap is less than $0.05E_0$ are excluded.
    }
    \label{fig:fcicriterea}
  \end{figure*}

We proceed to investigate the effects of $\kappa > 0$ on band energetics and band geometry.  The results are plotted in Fig.\ref{fig:fcicriterea} as a function of $\kappa$ and $\Delta \alpha = \alpha-\alpha_{\rm chiral}$, with $\alpha_{\rm chiral} = 0.568$ for the first magic angle \cite{Tarnopolsky2019}. The results of previous sections are valid for $\kappa = \Delta \alpha = 0$, and nonzero values of either lead to nonzero bandwidth and the violation of \eqref{trcond}.

We choose the (dimensionless) sublattice splitting $\delta = 0.10$, corresponding to $\Delta = 19$ meV \eqref{sublattice} for $w = 110$ meV. Realistic values depend on the hBN alignment angle and range from $17-30$ meV. Potentially we could consider $\delta$ as a tuning parameter. However this turns out to be redundant since larger sublattice potentials can be made to give similar results by tuning $\Delta \alpha$ as a function of $\kappa$ and $\delta$,  hence we do not pursue that direction here. Note that we consider the sublattice potential to be applied equally to both layers; in reality this is typically confined to a single layer adjacent to the aligned substrate. However, our results are similar to that obtained for a sublattice potential on a single layer if one simply doubles the strength $\delta \to 2\delta$.  
Finally, we neglect the rotation in kinetic terms and take $\vec \sigma_{\pm \theta/2} \to \vec \sigma$ in \eqref{continuummodel}. 

We quantify band energetics through inverse flatness ratios $R = \rm{Bandwidth}/\rm{Bandgap}$ to avoid divergences at $\kappa = 0$. There are bands above and below the band hosting the FCI state (the lower of the two flat bands) and so we have two different ratios: $R_{\rm upper}$ is the ratio of the bandwidth to the band gap to the upper band (upper flat band) and $R_{\rm lower}$ is the ratio of the bandwidth to the gap to the lower band (non-flat band).  Both ratios increase with increasing  $\kappa$, but can be lowered by judiciously {\em decreasing} $\alpha$ as $\kappa$ increases with $\Delta \alpha = \alpha - \alpha_{\rm chiral} \approx -0.02$. We do not consider points in parameter space where the bandgap to the lower band is less than $0.05 E_0$, as the near-crossing poses numerical issues and these points in parameter space are likely not suitable for FCI phases anyway.
 
  \begin{figure}
      \centering
      \includegraphics[width = 0.5\textwidth]{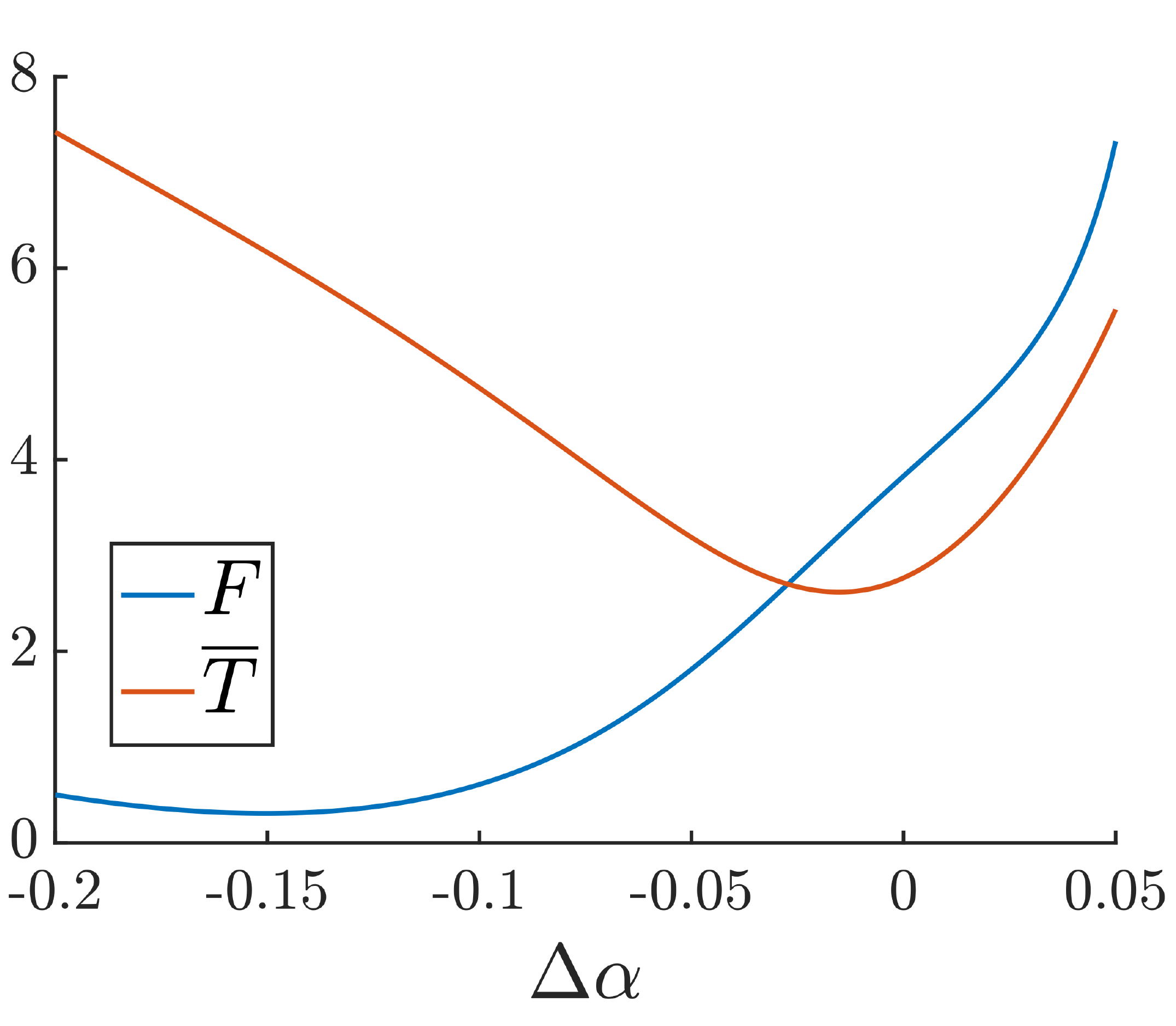}
      \caption{Berry curvature inhomogeneity (${F}$) and averaged violation ($\ov T$) of Eqn. \eqref{trcond} as a function of $\Delta \alpha$.  Berry curvature is most homogeneous at very negative $\Delta \alpha \approx -0.15$, whereas $\ov T$ is minimized for $\Delta \alpha \approx -0.02$. }
      \label{fig:alphadep}
  \end{figure}

\begin{figure}
      \centering
      \includegraphics[width=0.5\textwidth]{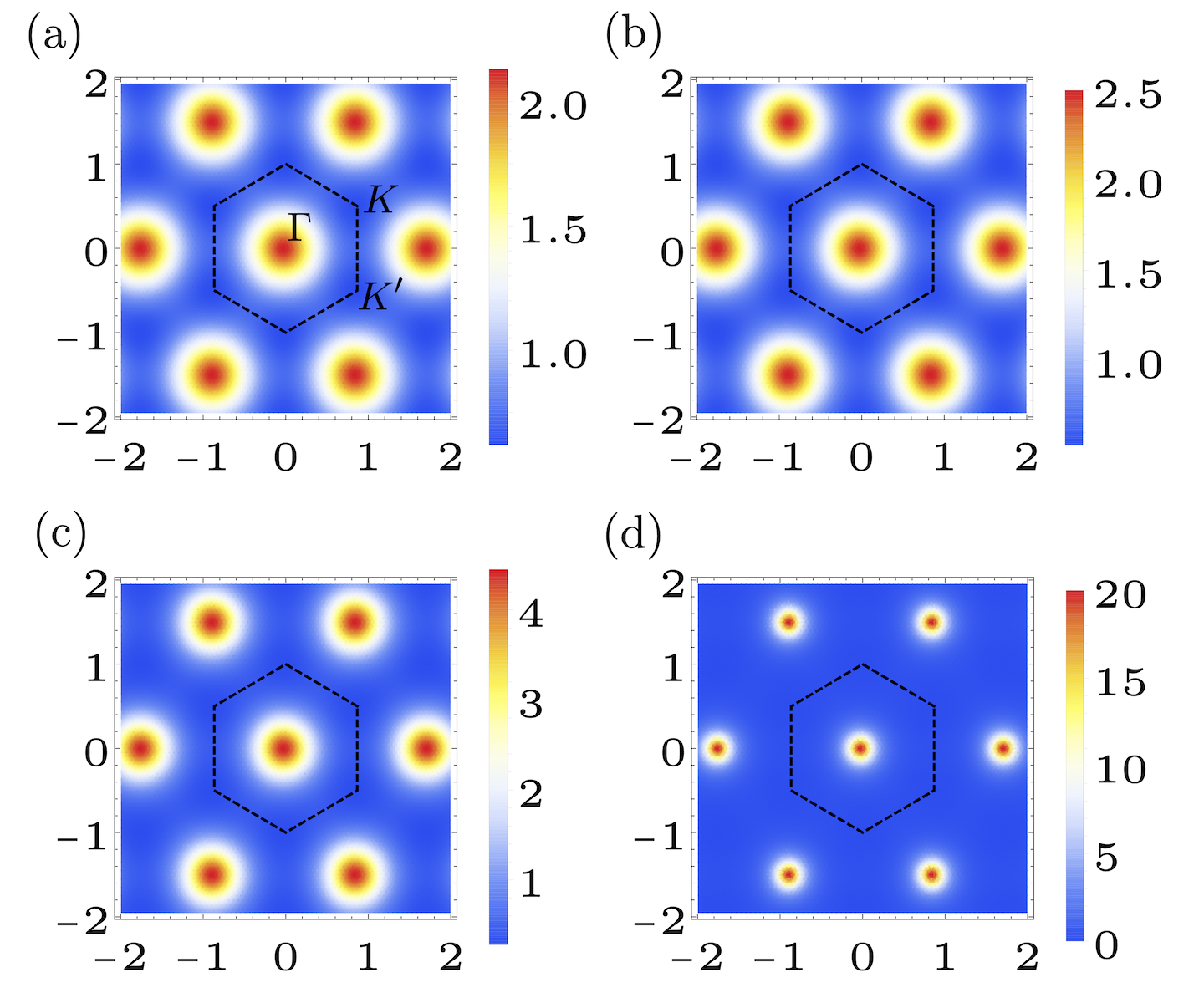}
      \caption{Berry curvature $\abs{\Omega(\bk)}$ as a function of momentum for the first magic angle at sublattice potential $\delta = 0.1$ for  $\kappa = 0, \, 0.25 \, , 0.5 \, , 0.75$ in (a) through (d) respectively.}
      \label{Fkappa}
  \end{figure}

  We now discuss the influence of $\kappa > 0$ on the wavefunctions and band geometry. We plot the distance from the chiral model wavefunctions, quantified by \begin{equation}
    d(\kappa) = 1-\langle \abs{ \braket{\psi_{\vec k}(\kappa)}{\psi_{\vec k}(0)} }^2 \rangle_{\rm BZ}.
    \label{dist}
  \end{equation}
  The distance increases with $\kappa$, and for larger $\kappa$ it decreases less when $\Delta \alpha$ is negative.  We also track the Berry curvature inhomogeneity $F$. For fixed $\alpha$, the Berry curvature distribution becomes less uniform as $\kappa$ is increased and begins to diverge as the band gap to the lower band closes. This is illustrated in Fig.~\ref{Fkappa} showing that the Berry curvature gets more concentrated at the $\Gamma$ point with increasing $\kappa$. 
As $\kappa$ increases $\ov T$ does too, though it does less so for negative $\Delta \alpha$.

One might expect that as one gets closer to $\kappa = 1$ the Bistritzer-Macdonald magic angle\cite{Bistritzer2011} would be more favorable than the chiral magic angle.  However, the Bistritzer-Macdonald angle corresponds to $\Delta \alpha = 0.02$ and we found above that \emph{negative} values of $\Delta \alpha$ are more suitable for FCI phases.  This counter-intuitive conclusion is important to take into account when choosing favorable parameters for future numerical and experimental studies. 

The dependence of $F$ and $\ov T$ on $\Delta \alpha$ is shown more extensively in Fig. \ref{fig:alphadep} for fixed $\kappa = 0.7$ and $\delta = 0.1$.  To minimize $\ov T$ one should choose $\Delta \alpha \approx -0.02$, but $F$ greatly decreases as $\alpha$ does and only reaches its minimum at $\Delta \alpha \approx -0.15$.  The values $\Delta \alpha = -0.15,-0.02,0,0.02$ correspond to the angles $\theta = 1.46^\circ, 1.12^\circ, 1.09^\circ, 1.05^\circ$ respectively, where we used $w = 110$ meV and $2vk_{D} = 19.81$ eV. The optimal value of $\Delta \alpha$ likely compromises between minimizing $\ov T$ and $F$, and perhaps optimizing band energetics, and probably lies between $-0.02$ and $-0.15$.

The values $F$ and $\ov T$ obtained for realistic $\kappa \approx 0.7-0.8$ are  not small.  However, they are often in the stable region of models in Ref.\onlinecite{Jackson2015}, although they are larger than the values of fine-tuned tight binding models in Table I.  For example, at $\kappa = 0.7$ and $\Delta \alpha = -0.075$ we have $F = 1$ and $\bar T = 4$ which is in the stable region of models in Ref. \onlinecite{Jackson2015}.  Still, since these parameters are not perturbatively small, it is important for future studies to numerically check how potential FCI states in fully realistic twisted bilayer graphene evolve on tuning away from chiral twisted bilayer graphene and the lowest Landau level. Our analysis can also be extended to other Moire systems where Chern bands appear as we discuss further below.

In the presence of the Coulomb interaction, interaction-driven band renormalization effects coming from normal ordering terms as well as the effect of remote bands become important \cite{XieMacDonald, Repellin19, Liu19, BultinckKhalaf, Guinea}. Due to the small "bare" dispersion, these effects alter the band dispersion significantly while keeping the flatband wavefunctions largely unchanged. Thus, we expect our conclusions in regard to the geometry of the flat band wavefunctions to be robust to such effects. On the other hand, we expect band dispersion to be strongly renormalized \cite{Liu19, BultinckKhalaf}. This is also supported by STM data \cite{CaltechSTM, ColombiaSTM, RutgersSTM, PrincetonSTM} which find a separation of around 10-15 meV between the van Hove peaks at empty filling which is significantly larger than the expected 3-5 meV based on the non-interacting band structure. We note, however, that despite the substantial bandwidth renormalization, it still remains a factor of 2-3 smaller than the main interaction scale \cite{Liu19, BultinckKhalaf}. This is also supported by STM data \cite{CaltechSTM, ColombiaSTM, RutgersSTM, PrincetonSTM} where the interaction-induced separation between the van Hove peaks at charge neutrality is significantly larger $\sim$35-50 meV. This suggests that the realization of an FCI is feasible in the realistic setting even when intraction induced bandwidth renormalization is taken into account.

  We also note that the interaction potential is a highly tunable quantity in a continuum model. This has been beautifully illustrated in two recent experimental works \cite{EfetovScreening, YoungScreening} which managed to tuned the interaction potential by changing the distance to the metallic gate which controls the screening of the interaction. The strong dependence of the obtained phase diagram \cite{EfetovScreening, YoungScreening} on the gate distance suggests the sensitivity of the different interaction parameters, including band renormalization effects, to screening. This makes it reasonable to expect the existence of some parameter regime where the FCI phases discussed here are stable.

\section{Conclusion}
Our results provide a good starting point for analyzing the possibility for FCI states in TBG through numerical tests of adiabatic continuity.  Indeed, one may start from LLL wavefunctions where the FQHE is well established, turn on the inhomogeneous part of the magnetic field until one reaches the chiral model wavefunctions, and then gradually turn on $\kappa$ until one reaches realistic wavefunctions.  If the many body gap does not close during this process, one has established adiabatic continuity.  One may also explore the continuous space of paths, parameterized by $\Delta\alpha$, $\delta$, and the strength and range of the interaction potential, and see if any lead to realizable FCI states.  

  Independent of relevance to current TBG experiments, our model is a promising idealized model for numerical studies of FCI stability.  Previous studies have learned much from toy lattice models\cite{Jackson2015} and Harper-Hofstadter models\cite{Bauer2016} but were restricted by lack of tunable parameters that produced distinct effects on band energetics and geometry.  Twisted bilayer graphene is highly tunable; it may be modified through (small) changes in angle, different strength sublattice splittings on each layer, a perpendicular electric field, pressure, screening length, and more.  
  It can add many more data points to these studies.

While the question of whether FCIs are hosted in realistic models of twisted bilayer graphene is best answered by numerical studies and experiments, we argue that our analytic work has the potential to be extended to provide answers to other questions not limited to twisted bilayer graphene or Laughlin-like FCI physics.  For example, our approach readily generalizes to Halperin states which could be especially important in twisted bilayer graphene as there is no Zeeman field to induce spin polarization.  Multicomponent textures like skyrmions also have a simple description in terms of first quantized holomorphic wavefunctions, and our work therefore enables these results in the lowest Landau level to be carried over to twisted bilayer graphene \cite{girvinreview}.  

Also, many other moir\'{e} systems exhibit flat bands with Chern number and can perhaps be well described by effective chiral models with holomorphic wavefunctions.  Our methods could be of use in predicting candidate phases for these systems as well (e.g. ``sandwich-stacked'' graphene \cite{CarrSandwich2019} and more generally alternating multilayers \cite{Khalaf2019}).

Finally, we argue that our methods have the potential to be generalized to other FCI systems.  Indeed, we have stressed the connections between the isotropic ideal droplet condition, holomorphic wavefunctions in real space, and the effective description in terms of a Dirac particle in an inhomogeneous magnetic field.  While generic models satisfying the isotropic ideal droplet condition do not immediately come with real-space holomorphic wavefunctions, tight-binding models certainly do not, we argue that it is very plausible that they admit an effective description in terms of a Dirac particle in an inhomogeneous magnetic field (such that all intraband matrix elements of density operators match).  Indeed, our model is not fine tuned and can be continuously deformed by changing $U(\vec r)$ without changing any of our qualitative conclusions.  Furthermore, all our conclusions are ultimately low energy and band-projected such that there is no fundamental distinction between continuum and tight-binding models.  We therefore argue that it is plausible that all models with ideal band geometry can be described in this way.  We hope future work will extend our conclusions and provide an answer to this conjecture.
  
{\bf Note Added:} After completion of this work, references \onlinecite{Ahmed2019,Repellin2019New} appeared, which contain numerical studies of FCIs in MATBG 
. Their  \onlinecite{Ahmed2019,Repellin2019New} observation of FCI states support our analytical results regarding the stability of Laughlin FCIs in realistic models of magic angle graphene and  \onlinecite{Repellin2019New} LLL-like physics at $\kappa = 0$.

{\bf Acknowledgements:} This work was partly supported by the Simons Collaboration on Ultra-Quantum Matter, which is a grant from
the Simons Foundation (651440, A. V.) and A. V., E. K., were supported by a Simons Investigator grant. E. K.
was supported by the German National Academy of Sciences
Leopoldina through grant LPDS 2018-02 Leopoldina fellowship. P.J.L. was partly supported by NSF-DMR 1411343 and the Department of Defense (DoD) through the National Defense Science \& Engineering Graduate Fellowship (NDSEG) Program. G.T.  acknowledges support from the MURI grant W911NF-14-1-0003 from ARO and by DOE grant de-sc0007870. 
\bibliography{sources}

\end{document}